\documentclass[preprint]{imsart}

\pdfoutput=1
\RequirePackage[OT1]{fontenc}
\RequirePackage{amsthm,amsmath,amssymb}
\RequirePackage{natbib}
\usepackage{graphicx}
\RequirePackage[colorlinks,citecolor=blue,urlcolor=blue]{hyperref}
\usepackage[margin=1in]{geometry}

\def\T{{ \mathrm{\scriptscriptstyle T} }}

\begin{document}	

\begin{frontmatter}
\title{Modelling discrete valued cross sectional time series with observation driven models}
\runtitle{Discrete Valued Cross Sectional Time Series}
\author{W. T. M. Dunsmuir \\School of Mathematics and Statistics, University of New South wales, Sydney, Australia.\\ W.Dunsmuir@unsw.edu.au \\
\and \\
C. McKendry \\Veda, 90 Alfred Street ,North Sydney, NSW 2060, Australia.\\Chris.McKendry@veda.com.au\\
\and \\
 R. T. Dean\\MARCS Institute, Western Sydney University, Sydney, Australia.\\ Roger.Dean@westernsydney.edu.au}
\maketitle

\begin{abstract}
	This paper develops computationally feasible methods for estimating random effects models in the context of regression modelling of multiple independent time series of discrete valued counts in which there is serial dependence. Given covariates, random effects and process history, the observed responses at each time in each series are independent and have an exponential family distribution. We develop maximum likelihood estimation of the mixed effects model using an observation driven generalized linear autoregressive moving average specification for the serial dependence in each series. The paper presents an easily implementable approach which uses existing single time series methods to handle the serial dependence structure in combination with adaptive Gaussian quadrature to approximate the integrals over the regression random effects required for the likelihood and its derivatives. The models and methods presented allow extension of existing mixed model procedures for count data by
	incorporating serial dependence which can differ in form and strength across the individual series. The structure of the model has some similarities
	to longitudinal data transition models with random effects. However, in
	contrast to that setting, where there are many cases and few to moderate
	observations per case, the time series setting has many observations per
	series and a few to moderate number of cross sectional time series. The method is illustrated on time series of binary responses to musical features obtained from a panel of listeners. 
	
\end{abstract}


\end{frontmatter}

\section{Introduction}

Increasingly, particularly in applications to public health,
epidemiology and biostastics, multiple independent time series of counts are
encountered. Often the main focus of analysis is assessment of regression
effects some of which are shared across series. As in typical longitudinal
data, the series corresponding to each case in the study will be assumed to be
independent of each other but there will be serial dependence and random
effects contributions to the correlation structure of the individual series.
In this regard the applications and models of this paper have a similar
structure to longitudinal data in which there are typically many cases each of
which have quite short observed time series.  The important
distinction here is that there are typically a small or moderate
number of medium to long time series.

Let $Y_{jt}$ be the observation at time $t=1,\ldots,n_{j}$ on the $j$-th
series of counts where $j=1,\ldots,J$. Given the regression variables, and any random effects, the
$Y_{jt}$ are assumed to be independent with a discrete response distribution. For longitudinal data examples $J$ is
typically very large compared with $n_{j}$. For example $J$ well might be in
the 100's or 1000's of cases or subjects while $n_{j}$ will be typically 5 to
10 and occasionally longer. \ In the time series examples we have encountered
$J$ is of the order of 5, 10, 20 or 50 but $n_{j}$ are large of orders such as 50, 100,
1000 for example. 

In all of these examples a major focus is on testing if regressors are
significant and if they have the same impact across the series. Some of these
regressors are common to all series while others are unique to individual
series. Another major focus is on characterizing the between subject variability in regression effects.
There is also the possibility that serial dependence exists within
each series and this needs to be accounted for in the regression analysis. Unlike the typical assumption that the form and structure of serial dependence is the same for each series the examples we have encountered require variation in the form and strength of serial dependence across the series.

\cite{DunsDVTS2016} reviews the obervation driven generalized linear moving average (GLARMA) models for single time series, gives a brief overview of how these may be extended to the long longitudinal (cross sectional time series) setting using both a fixed effects specification for regression coefficients and a mixed model specification which combines fixed effects and random effects for the regressors and applies the models and methods to Poisson counts of road deaths.   In the
fixed effects approach, models, which constrain the
regression and serial dependence parameters across series to reflect  hypotheses of interest, are fit using a likelihood obtained as the product of the likelihoods for individual
cases. In the random effects approach variations around fixed effects
parameters are modelled using random effects and the likelihood is constructed
by integration over the random effects. In this paper we explain the methodology for inference in the mixed model framework in detail and illustrate its use for an  analysis of listener binary arousal responses to acoustic intensity on a panel of $32$ listeners observed over $303$ time points. This example has a more complex random effects and serial dependence structure than that considered in \cite{DunsDVTS2016}.

There have been several recent contributions to the modelling of long longitudinal data and we briefly review these here. \cite{xu2007analysis} consider a model which combines serial dependence and random effects but not in the generality that is considered here. For example,
the random effect included in their model allows only for a serially
dependent random effect on the intercept term in the regression. In contrast, the
formulation presented here allows the traditional specification of correlated
random effects on several regression components. Additionally \cite{xu2007analysis} assume
that the serial dependence is of the same form and has the same parameter
value for all series being modelled. Such a restriction is not required in the
model presented here; indeed, the approach used here allows for that
assumption to be tested. This will be illustrated in the  example presented in Section \ref{Sec: Application}. In this and other data sets we have applied our model to the time series are considerably longer than
those considered in \cite{xu2007analysis}.

\cite{schildcrout2007marginalized} discuss moderate to long series of longitudinal binary response data using a marginalized model and with simple transition dynamics (and observation driven specification for serial dependence) and random effects (for long range temporal dependence). They apply their methods to data with 12 time points of observations. They allow the transition model parameters (autoregressive) and the random effects variance parameters to depend on covariates (dichotomized in the applications presented) something that we do not do.

\cite{Zhang2012} develop models specifically aimed at a small number of long sequences of count data. Their motivation is similar to ours. They point out that: ``Most of the available methods for longitudinal data analysis are designed and validated for the situation where the number of subjects is large and the number of observations per subject is relatively small." They use data collected from a panel of teenage drivers observed through time with low count responses. In this study there are 42 subjects (teenage drivers) observed for an average of 1626 trips unequally spaced in time. There are numerous covariates to be assessed. Serial dependence in each subjects observational trajectory is modelled by a continuous time autoregressive process with the same temporal correlation parameter across all subjects. They introduce a within-cluster resampling approach in conjunction with a marginal method of model estimation. Our methods would not apply directly to this setting because of the unequal temporal spacing but, if time were taken as observation number, they could provide a potential alternative modelling approach. An alternative description of serial dependence based on parameter driven models (see the concluding section) could be applicable to the unequally spaced time setting.

\cite{hung2008binary} propose a binary time series mixed model similar to that considered in this paper for application to adhesion frequency experiments. They introduce serial dependence through the BARMA type model of \cite{li1994time} which can be put in the form of the GLARMA model we use in this paper -- see \cite{dunsmuirglarma}. Their random effects are limited to an intercept random effect whereas we introduce multiple random effects in our treatment. Their estimation is based on an iterative procedure which first, given the intercept random effects variance, obtains estimates of the fixed effects, random effects and common serial dependence parameters using maximum partial quasi-partial likelihood and, secondly, uses the estimated random effects in a restricted maximum likelihood fit to get the random effects variance estimate, iterating between the two steps until convergence. In the method we propose the complete likelihood is calculated using adaptive Gaussian quadrature (AGQ) to approximate integration, with respect to the random effects distribution, of the conditional likelihood for each trajectory. Additionally we based optimization using analytical derivatives of the complete likelihood, evaluated using AGQ, to obtain estimates and their standard errors of all parameters in a single optimization. Our method does not require that all series have the same type or strength of serial dependence. This flexibility is not present in any other method we have encountered but is something that we have found necessary in all examples we have analysed to date.

\section{Models for multiple time series with random effects.}

As above, let $Y_{jt}$ be the observation at time $t=1,\ldots,n_{j}$ on the
$j$-th series of counts where $j=1,\ldots,J$ and let $x_{j,t}$ be the covariates for the $j$-th series.  We also let
$r_{jt}$ denote $d$ random effect covariates which apply to all series
through coefficients represented as vectors of normally distributed random effects
$U_{j}\sim\text{i.i.d\ }N(0,\Sigma(\lambda))$
where $\Sigma$ is a $d\times d$ covariance matrix determined by parameters
$\lambda$. In terms of these quantities we define a linear state process
\begin{equation}\label{Eq: Wjt general}
W_{j,t}=x_{j,t}^{\T}\beta^{(j)} +r_{jt}^{\T}U_{j}+\alpha_{jt}%
\end{equation}
for the $j$th case series where $x^\T$ denotes the transpose of a column vector. In addition to assuming that the series $\left\{  Y_{jt}\right\}  $ are
independent across the $J$ cases, for each $j$ we also assume that, given the state process
$\left\{W_{j,t}\right\}$,
$Y_{jt}$ are independent with exponential family distribution in the following form
\begin{equation}\label{Eq: ExpFamDensity}
f(y_{jt}|W_{jt})=\exp\left\{  y_{jt}W_{jt}-m_{jt}b(W_{jt})+c_{t}(y_{jt})\right\},
\end{equation}
where $\{m_{jt}\}$ are sequences of constants and $\{c_t(y_{jt})\}$ are defined in terms of the observed responses. 
For conditionally Poisson or binary data $m_{jt} =1$ for all $j$ and $t$ and for conditionally binomial observations $m_{jt} \ge 1$ can vary with $j$ and $t$. The conditional response distribution could also be negative binomial with an additional `shape' parameter but, for brevity, we will not give details for this case. Additional overdispersion could also be considered in the usual way.  

Throughout this paper we will focus
on the canonical link case with no additional dispersion parameter \cite{mccullagh1989generalized} for which the conditional mean is $\mu_{jt}=E(Y_{jt}|W_{j,t})=m_t\dot b(W_{j,t})$ and the conditional variance is $\sigma_{jt}^2=\textrm{Var}(Y_{jt}|W_{j,t})=m_t\ddot b(W_{j,t})$
where $\dot b$ and $\ddot b$  denote first and second derivatives with respect to the function argument. 

Serial dependence is modelled with the $\alpha_{jt}$ term in \eqref{Eq: Wjt general}. Two generic choices
for this process, referred to as \textit{parameter driven} and
\textit{observation driven} in \cite{cox1981statistical}, are reviewed in \cite{davis1999modeling}. For the
parameter driven specification, $\alpha_{jt}$ is a often taken to be a stationary Gaussian time series with covariance
function depending on a finite set of parameters. This specification is used
in Xu et al (2008). We will comment further on parameter driven versions of
the model at the conclusion of the paper.

This paper concentrates on the observation driven model in which $\alpha_{jt}$
is assumed to follow an autoregressive-moving average transition process
\citep{davis2003observation, DunsDVTS2016} with degrees $\left(  p^{(j)},q^{(j)}\right)  $%
\begin{equation}
\alpha_{jt}=\sum_{l=1}^{p^{(j)}}\phi_{l}^{(j)}\left(  \alpha_{j,t-l}%
+e_{j,t-l}\right)  +\sum_{l=1}^{q^{(j)}}\vartheta_{l}^{(j)}e_{j,t-l}%
\label{Eq: GLARMAgeneral}%
\end{equation}
In this paper we will use Pearson type innovations $e_{j,s}=(y_{j,s}-\mu_{j,s})/\sigma_{j,s}$ where $\mu_{j,s}$ and $\sigma_{j,s}$ are the conditional mean and standard deviation of $y_{j,s}$ defined above. These are zero mean, unit standard
deviation martingale differences with respect to the observed past. Other scaling for defining the residuals are discussed in \cite{dunsmuirglarma} and the methods presented in this paper would also apply to those. 

Let $\beta = (\beta^{(1)},\ldots,\beta^{(J)})$ denote the fixed effects regression parameters and $\tau=(\tau^{(1)},\ldots,\tau^{(J)})$, where $\tau^{(j)}=(\phi_1^{(j)},\ldots,\phi_{p^{(j)}}^{(j)},\vartheta_1^{(j)},\ldots,\vartheta_{q^{(j)}}^{(j)})$ denotes the serial dependence parameters for the $j$th series. Then $\theta
=(\beta,\tau)$ denotes the collection of all parameters needed to specify the state equations in \eqref{Eq: Wjt general}. To express various null hypotheses of interest we use constraints across these parameters: let $\psi$
be a lower dimensional vector of parameters in terms of which the regression coefficients and the serial dependence parameters across the $J$ series can be expressed. We limit these to linear contrasts and do not allow constraints between the regression and serial dependence parameters. Hence $\beta=A_{\beta}\psi_{\beta}$, $\tau=A_{\tau}\psi_{\tau}$ where $A_{\beta}$ and $A_{\tau}$ are constant matrices embodying the constraints. $\Psi=(\psi^{\T}, \lambda^{\T})$ is the complete collection of parameters needed to specify the above model. In our applications we will assume that the random effects covariance parameters $\lambda$ are the same for all series.

In order to use existing R language software \texttt{glarma} \citep{dunsmuirglarmapackage} to compute the likelihood and its derivatives for the above model at any value of the random effect, the parameters specifying the regression \textit{and} random effects components  of the state equation must enter linearly. To achieve this we parameterize
the covariance matrix as $\Sigma=LL^{\T} $ where $L$ is lower triangular and let $U_j=L\zeta_j$ where $\zeta_j$ are independent $N(0,I_d)$.
Let $\lambda=$vech$(L)$ denote half-vectorisation. With this re-parameterization $W_{jt}$ in \eqref{Eq: Wjt general} can be expressed linearly in terms of $\lambda$ as
\begin{equation}\label{Eq: WjtLinearBetaLambda}
W_{jt}  =x_{j,t}^{\T} \beta^{(j)}+\text{vech}(\zeta_{j}r_{j,t}^{\T} )^{\T} %
\lambda+\alpha_{jt}
\end{equation}
Note (\ref{Eq: WjtLinearBetaLambda}) is in the same
form as \eqref{Eq: Wjt general} but in which the parameters
$\lambda$ are treated as regression parameters for any fixed value of the
vector $\zeta_j$ and the random effects covariates $r_{j,t}^{\T}$.

There are covariance
structures for which the representation $\Sigma=LL^{\T}$ cannot be obtained such as when structural zeros in $\Sigma$ are required to reflect
hypotheses in which random effect components have zero covariance between them. However these can often
be accommodated by reordering the random effect variables and setting the
appropriate elements of $L$ to zero. In such cases $\lambda$ is the half
vectorization of $L$ with any structural zeros removed.

\section{Likelihood Estimation} \label{Sec: Likelihood Estimation}

The joint log-likelihood for the model specified by \eqref{Eq: ExpFamDensity}, \eqref{Eq: GLARMAgeneral} and \eqref{Eq: WjtLinearBetaLambda} is 
\begin{equation} \label{Eq: loglike full GenMod}
l(\Psi)=\sum_{j=1}^{J}l_{j}(\Psi)
\end{equation}
where
\begin{equation}
l_{j}(\Psi)=\log\int_{R^{d}}\exp
(l_{j}(\Psi|\zeta)g(\zeta)d\zeta\label{Eq: loglik j in z}
\end{equation}
in which $g(z)$ is the $\ d$-fold product of the standard normal density and
\[
l_{j}(\Psi|\zeta)=\sum_{t=1}^{n}\left[  y_{jt}%
W_{jt}-m_t b(W_{jt})\right] +\sum_{t=1}^{n}c(y_{jt})
\]
and $W_{jt}$ is defined in \eqref{Eq: WjtLinearBetaLambda}.

For any fixed $\Psi$ computation of the likelihood \eqref{Eq: loglike full GenMod} requires calculation of the $J$ integrals defined in \eqref{Eq: loglik j in z}. An approximate method based on the Laplace approximation and adaptive Gaussian quadrature is described in detail in Appendix A.
The integral in \eqref{Eq: loglik j in z} can be written as
\begin{equation} \label{Eq: Lj(Psi)}
L_{j}(\Psi)=\frac{1}%
{(2\pi)^{d/2}}\int_{R^{d}}\exp\left\{F_{j}(\zeta;\Psi)\right\}  d\zeta
\end{equation}
with
exponent for a particular series. This integral can be considered as a function of $\zeta$ for
fixed parameters $\Psi$,
\[
F_{j}(\zeta;\Psi)=\sum_{t=1}^{n}\left\{  y_{jt}W_{jt}(\zeta;\Psi
)-m_{jt}b(W_{jt}(\zeta;\Psi))\right\}  -\frac{\zeta^{\mathrm{\T} }\zeta}{2}%
+\sum_{t=1}^{n}c(y_{jt})
\]
where
\begin{equation}\label{Eq: WjtZform}
W_{jt}(\zeta;\Psi)=(r_{j,t}^{\T} L)\zeta+x_{j,t}^{\T} \beta^{(j)}+\alpha_{jt}.
\end{equation}
is treated as a function of $\zeta$ for the other terms fixed.
To find the Laplace approximation we expand the exponent $F_j(\zeta)$ around its
modal value in a second order Taylor series and integrate out the quadratic
term.

\subsection{Estimating the likelihood derivatives using adaptive Gaussian
	quadrature.}\label{Sec: LikeDerivsbyAGQ}

The first derivatives of the log-likelihood \eqref{Eq: loglike full GenMod} with respect to unknown parameters are
\begin{equation}\label{Eq: GLARMAlogLik jth 1stDeriv}
\dot{l}_{j}(\Psi)=\frac{\partial}{\partial\Psi}l_j(\theta)=\frac{1}{L_{j}(\Psi
	)}\int_{R^{d}}\frac{\partial}{\partial\theta}  l_{j}%
(\theta|\zeta)  \exp(l_{j}(\Psi|z)g(\zeta)d\zeta.
\end{equation}
from which the derivative of the overall likelihood is $\dot{l}(\Psi)=\sum_{j=1}^{J}\dot{l}_{j}(\Psi)$.
Second derivatives are
\begin{align} \label{Eq: GLARMAlogLik jth 2ndDeriv}
\ddot{l}_{j}(\Psi)  &  =\sum
_{j=1}^{J}\{\frac{1}{L_{j}(\Psi)}\int
\ddot l_{j}(\Psi|\zeta)  \exp
(l_{j}(\Psi|\zeta))g(\zeta)d\zeta \notag \\
&  +\frac{1}{L_{j}(\Psi)}\int_{\mathbb{R}^{d}}\dot l_{j}(  \Psi|\zeta)  \dot %
l_{j}[  (\Psi|\zeta )^{\T} \exp(l_{j}(\Psi|\zeta))g(\zeta)d\zeta \nonumber\\ & -\dot{l}_{j}(\Psi)\dot{l}_{j}(\Psi)^{\T}\}
\end{align}
which require order $S^2$ integrals to be approximated where $S$ is the length of $\Psi$.

The Laplace
approximation can provide quite accurate single point approximation to the
integrals required for the likelihood itself (being positive, unimodal and
with contours not too deviant from elliptical). However the first and second
derivative integrals have integrands that are certainly not positive, nor are
they unimodal and so a single point approximation should not be adequate nor is it when checked empirically. However adaptive Gaussian quadrature, can provide
multipoint approximations to the integrals in \eqref{Eq: GLARMAlogLik jth 1stDeriv} and \eqref{Eq: GLARMAlogLik jth 2ndDeriv}. The same quadrature points used in the AGQ approximation to \eqref{Eq: Lj(Psi)} are also be used for the derivative integrals in \eqref{Eq: GLARMAlogLik jth 1stDeriv} and \eqref{Eq: GLARMAlogLik jth 2ndDeriv}.

In our experience surprisingly few quadrature points are required to get
approximations to the likelihood and the first and second derivatives which are sufficiently accurate for convergence to the optimal
of the likelihood and which provide accurate standard errors for inference purposes -- see Appendix B and Section \ref{SubSec: Accuracy Times} or details. Since the same adapted quadrature points and weights are used
for estimating $l_{j}(\Psi)$, $\dot{l}_{j}(\Psi)$ and $\ddot{l}_{j}%
(\Psi)$, both the second and first order derivative approximations can be
computed in the same steps with the existing \texttt{glarma}
software. For speed of computation this is a real advantage. 

\subsection{Summary of Algorithm to approximate likelihood and derivative}

The alogorithm to approximate the likelihood and its derivatives from which optimization over the unknown parameters uses Newton-Raphson iteration. The strength of the proposed method is that it allows the existing and tested \texttt{glarma} software to be utilised in two ways. Firstly to find the modal point and associated Hessian required to determine the Laplace approximation and a set of quadrature points and, secondly, to compute the integrands at any combination o quadrature points and parameter values. Detailed derivations are provided in  Appendix A. We summarise the procedure as follows.

To optimize the AGQ estimated likelihood over $\Psi$t let $\Psi^{(k)}$ be the parameter value at the $k$th iterate:
\begin{enumerate}
	\item For each $j=1,\ldots,J$:
	\begin{itemize}
		\item Treating $\zeta$ as a parameter apply the \texttt{glarma} software to find the derivatives required for Newton-Raphson iteration to  obtain the mode $\zeta_{j}^{\ast}$ and Hessian $\Sigma_{j}^{\ast}$ of the integrand in \eqref{Eq: Lj(Psi)}. Find the Cholesky factor $K_{j}^{\ast}$ of $\Sigma_{j}^{\ast}$ and let $\zeta_{\mathcal{I}}$ denote the grid of $Q$ quadrature points in each of $d$ directions and $\mathcal{W}_{\mathcal{I}}$ the associated weights.
		\item Apply \texttt{glarma} software with $\Psi^{(k)}$ as the parameter value to calculate the integrands at the $d^Q$ integrating points $\zeta_{j}^{\ast}+K_{j}^{\ast
		}\sqrt{2}\zeta_{\mathbf{i}}$ for $\mathbf{i} \in \mathcal{I}$ to estimate $l_j(\Psi^{(k)})$ defined in \eqref{Eq: loglik j in z} and the derivatives $\dot{l}
		^{(Q)}(\Psi^{(k)})$, $\ddot{l}
		^{(Q)}(\Psi^{(k)})$ defined in \eqref{Eq: GLARMAlogLik jth 1stDeriv},  \eqref{Eq: GLARMAlogLik jth 2ndDeriv}.
	\end{itemize}
	\item Assemble complete likelihood and derivatives over $J$ cases.
	\item Use Newton-Raphson iteration to update $\Psi^{(k)} \to \Psi^{(k+1)}$ Repeat at step 1 until convergence.
\end{enumerate}

Let $S$ be the number of parameters in total in $\Psi$. In the case where the likelihood for each series component depends on all parameters in $\Psi$ there are
$J\times(1+S+2S(S+1)/2)=J(S+1)^{2}$, $d$-dimensional integrals to be calculated for optimizing the likelihood using the above algorithm. When each of the component series requires only some of the parameters in $\Psi$ the number of integrals reduces slightly. 

\subsection{Large Sample Distribution}

For inference purposes 
as $J$ and $n_1 \ldots, n_J$ increase we assume that $\hat \Psi \to \Psi$ almost surely and 
\begin{equation} \label{Eq: CLT}
\sqrt{Jn}(\hat \Psi -\Psi) \overset{d}{\to} \textrm{N}(0,\Omega(\Psi)) \quad \textrm{where} \quad \Omega(\Psi) = \lim_{J,n \to \infty} \frac{1}{Jn}\ddot{l}_{j}(\Psi).
\end{equation}
Using the large sample distribution in \eqref{Eq: CLT}, approximate tests of significance for individual coefficients can be performed in the usual way as can Wald tests for several coefficients. Similarly the likelihood ratio test can be used with the chi-squared distribution used as an approximate sampling distribution. However, as in any tests of parameters in which the variance of the random effects is zero under the null hypothesis the usual chi-squared distribution will not apply without modification to the significance level - see \cite{fitzmaurice2012applied} for example.

In the typical longitudinal setup $J \to \infty$ and each $n_j$ is finite and it is assumed that parameters describing the regression fixed effects and the serial dependence are such that they are common across subsets of series each of which increases as order $J$.

In the long longitudinal set up we conceive of the number of terms in each series $n_j \to \infty$ without necessarily requiring $J \to \infty$. In these situations the serial dependence parameters can be unique to each series as can some or all of the fixed effects regression parameters. However, there is very little theory available for even the single time series asymptotic distribution for observation driven models -- see \cite{DunsDVTS2016} for a recent review. Rigorous results about consistency and asymptotic normality are only available in very special cases of scalar series and also have not been extended to the multiple independent series case considered here. Hence, at this stage, the central limit theorem of \eqref{Eq: CLT} is used heuristically to assess significance of coefficients and associated test statistics. 

\section{Application to Binary Panel Time Series of Music Listeners} \label{Sec: Application}
\subsection{Background}
This application arises in studies of time series of real-time responses by a panel of listeners to musical features as described in \cite{DeanBailesDunsmuirJMMpt1}. Details of the experiments, musical selections and methods of measuring listener response are given there and references therein. Here we provide brief details only. In this particular example, one of many such experimental data sets obtained by Dean and his collaborators \citep{DeanetalPLOS1}, there is a panel of $32$ listeners comprising $8$ electro-acoustic musicians (EA, listeners 1:8), $8$ musicians (M, listeners 9:16) and 16 non-musicians (NM, listeners 17:32) each of which provides continuously recorded responses to the same segment of music. In studies of this type interest is in quantifying the relationship between musical features, here acoustic intensity, and listener responses and examining the extent to which these differ between individuals and between musical expertise. The main response analysed was listeners' continuously recorded perception of the arousal expressed by the music.

\begin{figure}
	\centering
	\makebox{\includegraphics[height=4.2774in, width=2.9715in
		]{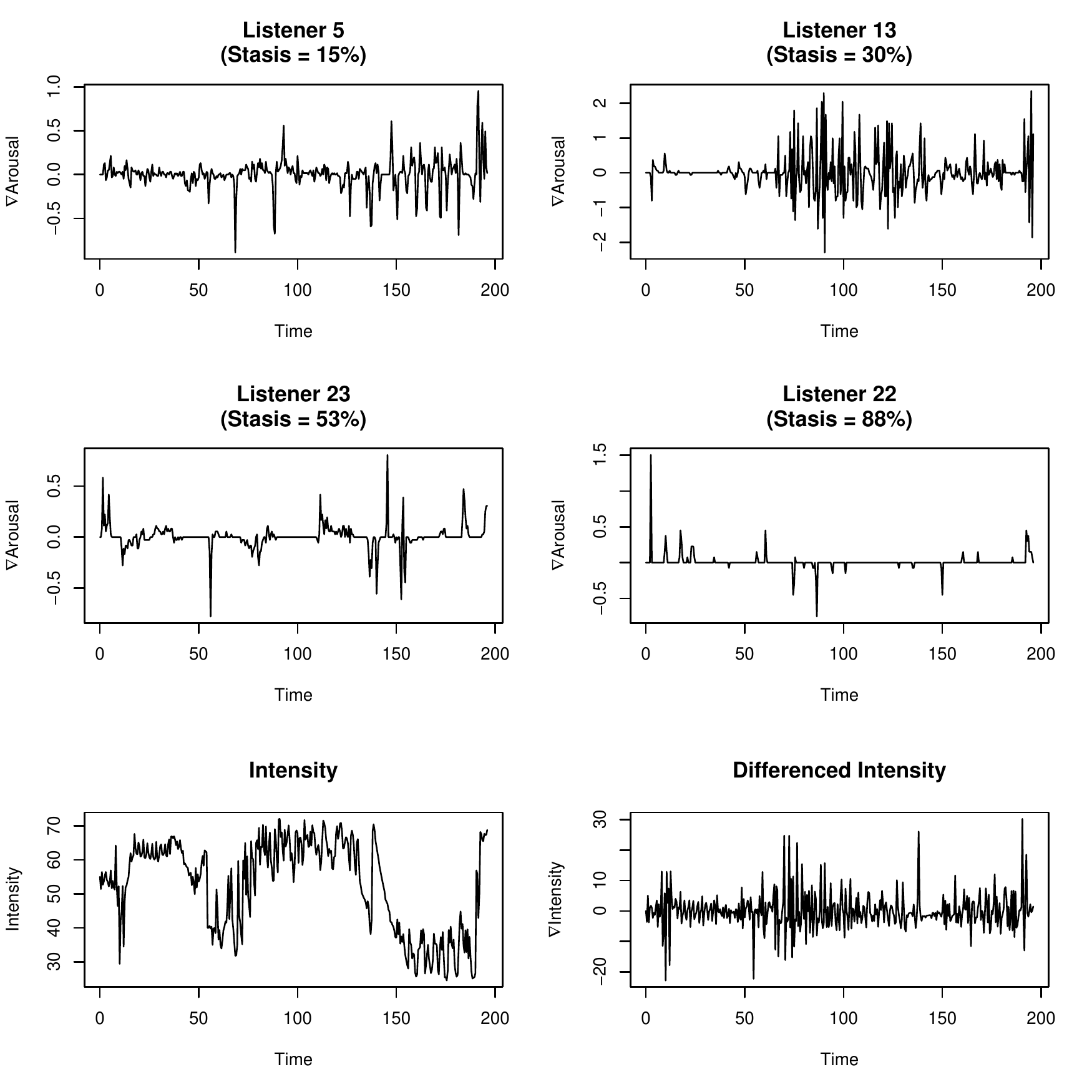}}
	\caption{\label{Fig: NablaArousalTSwithNablaIntensity_listeners5_13_23_22} Lag 1 differenced standardised arousal series for two listeners. (Percentage of
		times that the difference is zero is shown as  $\%$ Stasis) and differenced acoustic intensity.}
\end{figure}

Changes in arousal responses $\nabla A_{j,t}=A_{j,t-1}-A_{j,t}$ for four selected listeners are shown in Figure \ref{Fig: NablaArousalTSwithNablaIntensity_listeners5_13_23_22} along with acoustic intensity $I_t$ at time $t$ and its lag difference, $\nabla I_t$, for the extract from Wishart's Red Bird electro-acoustic composition.
These were selected to illustrate the wide proportion of stasis (when a response does not change from the previous response) occurring across the 32 series. Responses for listeners 5 and 13, while having a substantial percentage of zero
values may be amenable to modelling using standard continuous time series
modelling. However
responses for listeners 23 and 22 have substantial proportions of zero
differences and for these continuous distribution assumption models
are clearly inappropriate.

To assess the impact of substantial amounts of stasis on the robustness of their conclusions \cite{DeanBailesDunsmuirJMMpt1} consider the binary series 
$Y_{j,t}=0$ if $\nabla A_{j,t}\leq0$ and $Y_{j,t}=1$ if $\nabla A_{j,t} > 0$. Given the past observations on each series $Y_{j,t}=0$ is a binary response with density \eqref{Eq: ExpFamDensity} where $m_t \equiv 1$, $\mu_{j,t}$ is the conditional probability of
a positive change occurring at time $t$ for series $j$ and the log odds of a positive change relative to a negative or no change is $W_{j,t}=\mathrm{logit}(\mu_{k,t})$ where
\begin{equation}\label{Eqn:TF11State}
W_{j,t}  =\beta_{j,0}+\sum_{l=1}^{11}\omega_{j,l}\nabla I_{t-l}+\alpha_{j,t}
\end{equation}
and $\alpha_{j,t}$ is modelled by \eqref{Eq: GLARMAgeneral} using Pearson residuals. An autoregressive process with $p_{j}=1$ and $q_{j}=0$ was adequate to model the serial
dependence in all $32$ binary time series giving $\alpha_{j,t}$ as a simple geometrically decaying weighted average of past residuals.
\begin{equation}\label{Eq:GLARerror}
\alpha_{j,t}=\sum_{l=1}^{\infty}\phi_{j}^{l}e_{j,t-l}.
\end{equation}
Since $\hat \phi_{j}>0$ was observed for all 32 individual series there is positive serial dependence in each
state variable $W_{j,t}$. Since $e_{j,t}$ are zero mean, unit variance martingale differences with respect to $\mathcal{F}_{j,t-1}$, the unconditional expected value of the logit of the probability of $Y_{j,t}$ is $E(W_{j,t})=\beta_{j,0}+\sum_{l=1}^{11}\omega_{j,l}\nabla I_{t-l}$.

\begin{figure}
	\centering
	\makebox{\includegraphics[
		height=4.2774in,
		width=2.9715in
		]{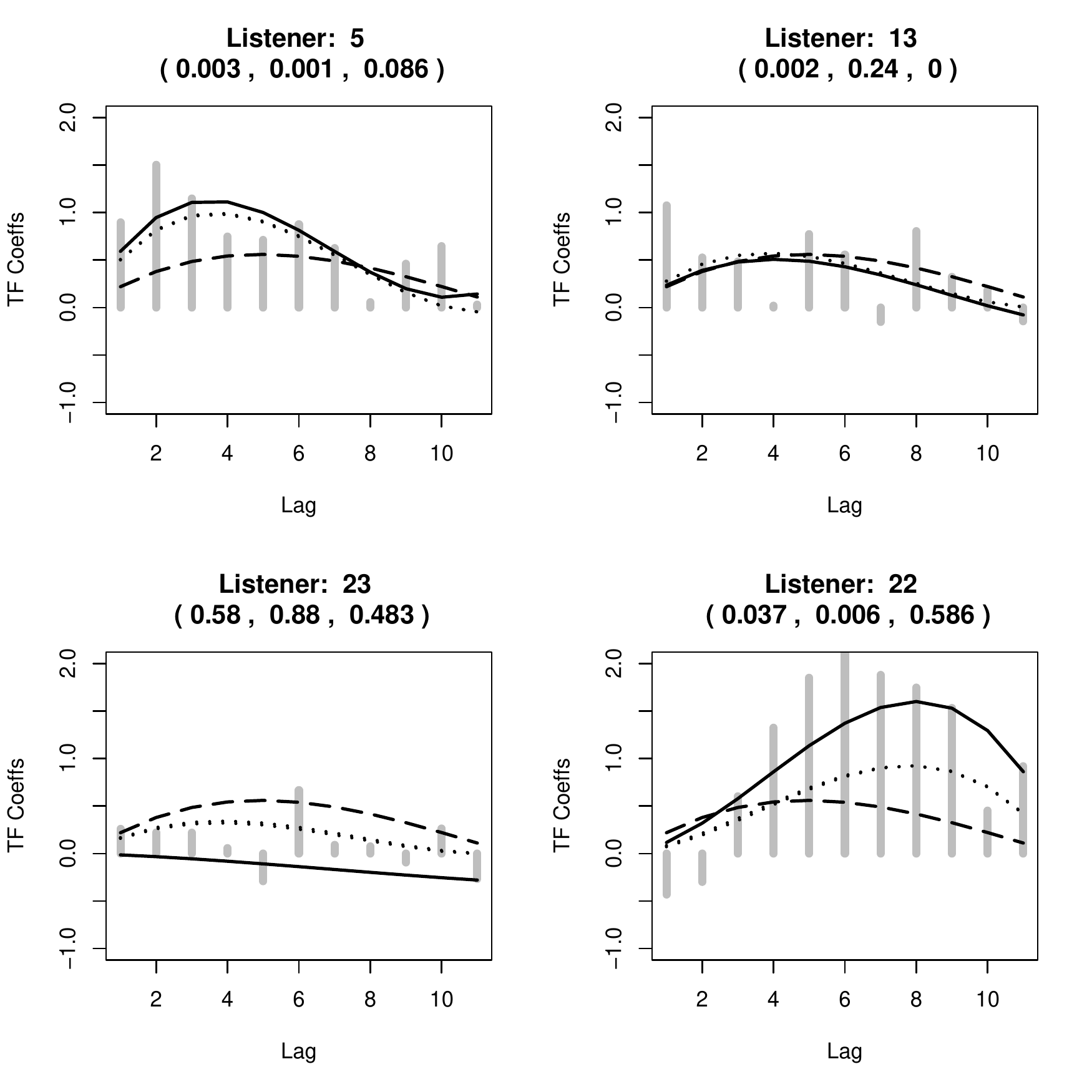}}
	\caption{\label{Fig: CubFitwithREfitCases5_13_23_22} Fitted transfer functions coefficients: unstructured 11 lag individual transfer function coefficients: gray vertical bars; orthogonal cubic model individual fits: solid line; fixed effect orthogonal cubic component of mixed model: dashed line; posterior mean of transfer function estimate from mixed model: dotted line. Numbers shown in the panel titles correspond to P-values for fit of unstructured TF, cubic TF and significance of unstructured TF to cubic in the individual fixed effects GLARMA models.}
	
\end{figure}

Figure \ref{Fig: CubFitwithREfitCases5_13_23_22} shows (as vertical grey bars) the individual estimates of the transfer function coefficients  $\omega_{j,1},\ldots,\omega_{j,11}$ in \eqref{Eqn:TF11State} for the same four selected listeners. The first value in parenthesis is the $P$-value for significance of the transfer function overall. This was significantly different from the zero transfer function ($\omega_{j,11}=0, j=1,\ldots,11$) for three of the four listeners and non significant for listener 23. \cite{DeanBailesDunsmuirJMMpt1} applied this approach to all $32$ binary series and the transfer functions compared to those using the continuous response ARMAX model. Overall they concluded that the binary analysis was supportive of the conclusions obtained from the traditional transfer function approach. In \cite{DeanBailesDunsmuirJMMpt2} random effects modelling was applied to the panel of all 32 responses again with the caveat that this method assumes continuous valued responses. In Section \ref{Sec: RE model} we apply the the random effects GLARMA model specified by \eqref{Eq: Wjt general}, \eqref{Eq: GLARMAgeneral} to the collection of binary responses $Y_{j.t}$.

\subsection{Simplifying the transfer function lag structure}

The random effects component of \eqref{Eq: Wjt general} has to be specified. If random effects were allowed for the intercept term and the  $L=11$ transfer function cofficients in \eqref{Eqn:TF11State} this would require $12$ random effects. This is not feasible with our method nor is it with current methods for non-linear mixed model software in general.  Plots of the unstructured transfer functions for all $32$ series suggested that their shape might well be modelled using a reduced set of a few simpler functions such as polynomials or splines. Additionally, four and five principal components accounted for approximately $92\%$ to $95\%$ of the variation of the unstructured transfer function coefficients between subjects suggesting that these coefficients could be represented by $4$ to $5$ basis functions of which one is the overall mean level. 

For this analysis the individual transfer functions were represented parsimoniously in the following way. Let $h_k: [0,1] \to \mathbb{R}$ for $k=1, \ldots, K$ be specified functions and let $H_k(l) =h_k(l/12) $ for $l=1,\ldots,11$ and $k=1,\ldots,K$.  Define new regressors as
\begin{equation}\label{Eqn: PolyRegors}
X_t^{(r)} = \sum_{l=1}^{11} H_s(l)\nabla I_{t-l},r=1,\ldots,K
\end{equation}
In effect this represents the $11$-lag transfer functions coefficients in (\ref{Eqn:TF11State}) as
\begin{equation}\label{}
\omega_l = \sum_{k=1}^{K}\beta_k H_r(l),\quad l=1,\ldots,11
\end{equation}
in terms of new parameters $\beta_s$.
to re-express (\ref{Eqn:TF11State}) as
\begin{equation}\label{Eq:StateEqnParm}
W_{j,t} =\beta_{j,0}+ \sum_{r=1}^{K} \beta_{j,k}X_t^{(r)}+\alpha_{j,t},\quad j=1,\ldots,32.
\end{equation}
For the analysis presented here the polynomials were defined as: 
$h_1(v)=v$, $h_2(v)=v(1-v)$,  $h_3(v)=(1-2v)h_2(v)$ and $h_4(v)=((1-14/3*v+14/3*v^2))h_2(v)$.
These are orthogonal over $[0,1]$ and satisfy zero end conditions $h_j(0)=0$ for $j=1,\ldots,4$ and $h_j(1)=0$ for $j=2,\ldots,4$.
This choice balanced parsimony of regression and random effects specification with individual series transfer function behaviour.

We fit the above binary time series model with cubic transfer function to each individual
listener's binary response series $Y_{j,t}$. The fitted transfer functions using the cubic model were overlaid with the unstructured transfer functions for all $32$ series. Figure \ref{Fig: CubFitwithREfitCases5_13_23_22} shows, for the four example series, the cubic fits (as the solid line in each panel) overlaid with the $11$ lag unstructured transfer function. Also shown in this figure are the $P$-values for testing that the transfer functions are significant or differ significantly. The third of the three $P$-values suggests that on an individual basis the cubic model is a sufficient summary of the complexity in the unstructured $11$ lag transfer function except for listener 13 in which the spike in the transfer function at lag 1 is not well capture by the smooth cubic polynomial. Overall, the cubic was sufficiently flexible to capture the key lag response features for almost all series. In two or three instances if there was a significant difference it was due to a one or two isolated lags. Higher order polynomial terms (such as quartic) were not beneficial in improving fit to the individual listeners' transfer functions. The $32$ individual autoregressive parameter estimates $\hat \phi_j$ could be simplified to one of $5$ values.

\subsection{The Random Effects Model} \label{Sec: RE model}

We now consider the general model \eqref{Eq: Wjt general} with both fixed and random effects covariates given by an intercept and $X_t^{(k)}$ for $k=1,2,3$ defined in \eqref{Eqn: PolyRegors} using the constrained orthogonal linear, quadratic and cubic functions defined previously.

Analysis of the estimates of $\beta$ for the $32$ individual listeners described in the previous section indicated that the intercept, quadratic and cubic coefficients had substantial variance across subjects while the linear term did not. Additionally, based on the correlation matrics an initial specification of the covariance matrix $G$ placed zero correlation on the intercept with linear and quadratic terms and between the quadratic and cubic term resulting in 7 parameters in $\lambda$. Comparing this model with that in which random effect on the linear term (and the covariances between it and cubic and quadratic terms) was not included gave a likelihood ratio statistic of $G^2 = 0.67$ on $3$ degrees of freedom) which is not significant, even accounting for the fact that this is a test of a parameter on the boundary as in \cite{fitzmaurice2012applied}. The resulting random effects structure could not be simplified further.  We also refit this model but allowed the four fixed effects regression parameters to vary between musician group (a hypothesis of interest in this study). The null hypothesis that the fixed effects do not vary between groups was not rejected ($G^2 = 9.69$ on $8$ degrees of freedom). Hence, in the model in which between listener variation is accounted for using random effects significant differences between groups was not detected confirming the conclusion based on individual fixed effects model of the last section. In the model with the same fixed effects across musician groups the linear fixed effects term was not significant ($\hat \beta_1= 0.078$, standard error $0.093$). Note, that the cubic fixed effect term in the model is also not significant. However, it was retained because the random effect for the cubic term was significant.

The final model has quadratic and cubic orthogonal regressors 
\begin{eqnarray} \label{Eqn: Final Model Wjt}
W_{j,t} = (\beta_0 + U_1) + (\beta_2 +U_2) X_t^{(2)} + (\beta_3+U_3) X_t^{(3)}+\alpha_{j,t}
\end{eqnarray}
where $\alpha_{j,t}$ is as in \eqref{Eq:GLARerror} and the value of the serial dependence parameter $\phi_j$ takes $5$ different values corresponding to the following grouping of listeners: (24, 25),
(5, 13, 20, 21),
(1,  8,  9, 10, 11, 15, 19, 22, 26, 29, 31),
( 2,  4,  6, 7, 12, 14, 16, 27, 30, 32),
( 3, 17, 18, 23, 28).
Parameter estimates and their standard errors are summarized in Table \ref{Tab: RE Estimates}. Note that $\Sigma(\lambda)=LL^{\T}$ gives random effects variance of the intercept term as $0.74$, quadratic term as $2.08$, the cubic term as $7.97$ and correlation between the intercept and cubic terms as $0.54$.


The posterior mean estimates of transfer functions were approximated by integrals using the same quadrature points as used to compute the likelihood and are shown in Fig. \ref{Fig: CubFitwithREfitCases5_13_23_22} for the four illustrative examples. As expected these posterior mean transfer function curves lie between the individual fixed effect fit and the fixed effects part of the transfer function calculated using the fixed effects component of the random effects model just described -- note, while this is the same for all listeners it is not the marginal mean over the panel. Note that listener 23 is an example of someone who expresses no significant response to the acoustic intensity in this particular musical extract.

\section{Computational Aspects}\label{Sec: Computational Aspects}

\subsection{Optimizing the Likelihood Approximation.}

Optimisation of the approximation to the likelihood $l^{(Q)}(\Psi)$ was attempted using the  \texttt{optim} function in R without providing derivative information. 
This approach proved to be extremely slow and often failed to converge to the global optimum value even from reasonable starting values. These approaches might be improved by providing functions to calculate analytical
derivatives. However these require implicit differentiation of $\zeta_{j}^{\ast}(\Psi)$ and
$\Sigma_{j}^{\ast}(\Psi)$
similar to \cite{huber2004estimation}. Expressions for the implicit derivatives were derived but they have the disadvantage that they cannot be
calculated using the existing single series GLARMA software and require
considerable additional programming effort to implement. The automatic differentiation approach of \cite{skaug2006automatic} which as been applied to single series parameter driven models for Poisson counts would also require substantial investment in software development. An alternative is to use optimization methods based on numerical derivatives obtained using the R package \texttt{numDeriv}. These also proved to be at least 200 times slower than the use of the Newton-Raphson method based on approximations to the analytical derivatives that we describe in Section \ref{Sec: LikeDerivsbyAGQ}.

\subsection{Accuracy and Execution Time Tradeoff} \label{SubSec: Accuracy Times}

For the final model of the Section \ref{Sec: RE model} specified by \eqref{Eqn: Final Model Wjt} there are $S=12$ parameters resulting in 2592 integrals to be estimated for calculation of the likelihood and its first and second derivatives with respect to $\Psi$, each integral requiring $3^Q$ quadrature points. The On-line supplement gives comparison of accuracy and computational time trade-offs for varying values of the number of quadrature points $Q$. Sufficient accuracy for likelihood optimization and standard errors using $Q=5$ and this is done in resonable computational time. Smaller values of $Q$ can be used in the early iterations for optimizing the likelihood. 

These conclusions are in line with those observed in application of the algorithm proposed in this paper to other examples. For the road deaths Poisson response example reported in \cite{DunsDVTS2016} with $J=17$ series each containing $n_j = 72$ monthly observations and using $S= 10$ model parameters and $d=2$ uncorrelated random effects, the total number of integrals required was $1377$. Comparing $Q = 3$ with $Q = 5$ the parameter estimates differed occasionally in the 4th decimal place at most, $-2ll$ in the 3rd decimal place and standard errors of parameter estimates in at most the 3rd decimal place. When the number of quadrature points was increased to Q = 7 there was no difference to the fourth decimal place in estimates or standard errors compared with the Q = 5 results. 
Computational time in this example was of the order of seconds compared with minutes shown in Table \ref{Tab: TimeAccuracy}. This is largely due to the length of the series being about $25\%$ of that in the example of this paper.

In another example using a Poisson distribution to model $n_j=336$ monthly suicide counts in $J=35$ US states with $S=13$ parameters, of which $7$ were fixed effects, $2$ were serial dependence parameters and the $d=3$ random effects required $4$ parameters, $Q=1$ (Laplace approximation) gave very accurate estimates of the log likelihood, $Q=5$ very accurate estimates of the first derivative vector and sufficiently accurate estimates of the Hessian. Use of  $Q=7$ ensured the Hessian was very accurately estimated.

Computation time could be substantially reduced using multiprocessor technology. To compute the Laplace approximation and quadrature points required to estimate the likelihood and its derivatives at any parameter value $\Psi$ the contributions for each of the $J$ cases can be done in parallel on separate processors. The code currently uses vectorized arithmetic to compute the  integrands at the collection of quadrature points for each case. Parallelisation of this should allow the number of cases to increase without substantial increase in computation time. However, because recursive calculation (in time index $t$)  is required to compute the GLARMA model likelihood (once for Laplace approximation and again at each quadrature point) increase in series lengths $n_j$ will result in slower overall computational costs which cannot be improved by parallel processing.

\section{Discussion}

This paper has presented in detail a computationally feasible methodology for modelling long longitudinal data sets allowing for shared and individual fixed effects parameters, random effects across series and individual serial dependence structure and strength across series. Serial dependence is modelled using the observation driven GLARMA class of models and this allows rapid computation of the likelihood contribution from each series. Serial dependence can also be modelled using a parameter driven or latent process specification but this requires estimation of extremely large integrals for each series in addition to those required to handle the random effects components of the model. 

From the point of view of the musical analyses which stimulated some of this work, the results here generally strengthen previous interpretations, and in particular place those involving long periods of stasis on a more robust basis. The models show clearly that aspects of the lag structure of the transfer function from acoustic intensity to perceived arousal vary considerably amongst individuals, and indeed they may be grouped on this basis. However, the resultant groupings do not coincide with the expertise groupings which constituted the participant pool, that is EA, M and NM. Indeed, when the random effects structure is reasonably modeled, using the cubic transfer function, differences between the expertise groups are not significant. Nevertheless, as shown previously in simpler models, group effects can be demonstrated. Thus overall, it seems, as suggested in the previous work, that it is the magnitude of inter-individual variation which dominates over the group effects.  Conceivably with larger groups of participants, it might be possible to quantitate the smaller contribution of expertise group alongside the larger effects of individual variation. 
	
Figure 3 reveals the complexity of individual responses to intensity, and encourages caution in going beyond suggesting such a possibility. Nevertheles, it is interesting that each of the five groups of participants identified on the basis of their strength of serial dependence  tend not to be dominated by a single musical expertise group. This suggests that there are features of this electro-acoustic piece, Red Bird, which are impinging in a similar way amongst a group, but that recognition of these features is not expertise dependent.  \cite{Bailesetal2012} showed earlier that all but a few members of the EA group were unfamiliar with this particular piece, and only the EA group had familiarity with the broad musical genre of which it is a member, so perhaps it is not surprising that the M and NM participants might contribute to each of the 4 main TF groups. One perhaps optimistic interpretation is that the  features driving perception of arousal, beyond intensity, may be so transparent or dominant that each group can respond to them, even without familiarity. In earlier work we also showed the power of certain animate and human sounds in the piece in predicting aspects of the continuous response: these may be amongst such relevant features. 

This paper has focussed on the binary response case. 
This is one element of a complete decomposition model of the type that has been successfully applied in
modelling high frequency financial data. For example \cite{rydberg2003dynamics} decompose series of stock transactions into a binary process for a change in price or not, a second binary process for a positive or negative change when one occurs and a count process for the size of price change in ticks. An alternative is a trinomial process for the occurrence of a negative change, no change or a positive change and a component for the size of change as suggested in \cite{liesenfeld2006modelling}.  The analogous model in our setting would decompose the series of lag 1 differences in arousal response into a binary process for a change in arousal, a second one for the sign of the change (positive or negative) when one occurred and a third component to denote the size of change when one occurs.

\cite{DunsDVTS2016} applied the models and methods of this paper to the Poisson response case. It is a straightforward matter to extend the methods and ideas exposited here to binomial
and negative binomial response distributions.

As mentioned in the introduction, \cite{xu2007analysis} develop a parameter driven model with random effects. We are currently developing an analogous approach using parameter driven models for the serial dependence. \cite{davis2005estimation} have successfully developed rapid and accurate approximate importance sampling methods based on the Laplace approximation for a single time series and our aim is to combine that methodology with the adaptive Gaussian quadrature methodology for derivatives proposed in this paper.

For some applications the assumption of independence
between series used in this paper is overly simple. In spatio-temporal settings there is a need
to link the series with a spatial correlation structure. Very
general models of this type in which the $\alpha_{jt}$ series are unobservable
latent stationary random processes are considered in \cite{kuo2008quasi} but
those models require evaluation of very large integrals to compute the
likelihood. For the observation driven models considered
in this paper a new formulation of the $W_{j,t}$ process is required to include
spatial correlation and random effects. It would be desirable to include these
terms in a way that preserves the simple approach to estimation described
above in order that models can be applied in practice relatively easily.

Given the variation in the autoregressive and moving average coefficients of the GLARMA model \eqref{Eq: GLARMAgeneral} that has been observed in practice and consistent with modelling variation between individual series regression coefficients, it could be reasonable to also use a random effects specification on these serial dependence parameters. In the above modelling of binary responses we grouped the autoregressive coefficients $\phi_j$ into five groups with a separate parameter representing series in each group. A random effects specification could be used for the serial dependence parameters but this would require a transformation from unrestricted real space for reparameterized partial autocorrelations to autoregressive coefficients as used in \cite{jones1987randomly}. This approach is under development.

In summary we have presented a class of models for cross sectional time series which combines the computational advantages of observation driven models for serial dependence in each series with adaptive Gaussian quadrature for integration over the random effects distribution. The number of random effects that can be handled using the methods presented can be of the same order as in standard mixed effects modelling packages and, particularly important, individual series can have different specifications of serial dependence structure and strength. We know of no other models and methods for discrete valued cross sectional time series data that have this necessary flexibility. A major conclusion of the analysis presented in this paper is that adaptive Gausssian quadrature integration designed to approximate the likelihood can also be used to obtain approximate analytic first and second derivatives of it with respect to the parameters. These approximations are sufficiently accurate for relative low numbers of quadrature points and hence lead to acceptably fast computation time, likelihood optimization and accurate standard errors. Speed comparisons with standard built in optimizing functions or numerical derivatives suggest the methods proposed here are as accurate with less than one two hundredth of the computational effort. This speed advantage means that models of the type considered here can be fit relatively quickly on modestly powered computers. Additionally the computations are amenable to parallelization with respect to the number of series which would allow application of these models and methods to cross sectional settings with a much larger number of time series.
\section{Appendix A: Laplace approximation and adaptive Gaussian quadrature}

To find the Laplace approximation to the integral 
\begin{equation} \label{Eq: Lj(Psi) Supplement}
	L_{j}(\Psi)=\frac{1}%
	{(2\pi)^{d/2}}\int_{R^{d}}\exp\left\{F_{j}(\zeta;\Psi)\right\}  d\zeta
\end{equation}
we expand the exponent $F_j(\zeta)$ around its
modal value $\zeta^{*}_j$ in a second order Taylor series and integrate out the quadratic
term. The derivatives required for this are
\[
\frac{\partial}{\partial \zeta}F_{j}(\zeta)=\sum_{t=1}^{n}\frac{\partial}{\partial
	\zeta}\left[  y_{jt}W_{jt}(\zeta)-b(W_{jt}(\zeta))\right]  -\zeta
\]
\[
\Sigma_j(\zeta)=\frac{\partial^{2}}{\partial \zeta\partial \zeta^{\T}}F_{j}(\zeta)=\sum_{t=1}^{n}\frac{\partial^{2}%
}{\partial \zeta\partial \zeta^{\T}}\left[  y_{jt}W_{jt}(\zeta)-b(W_{jt}(\zeta))\right]  -I.
\]
in which the $W_{j,t}$ terms and their derivatives require recursive calculation. These can be obtained using the existing \texttt{glarma} software by setting $x_{j,t}^{\T} %
\beta^{(j)}$ as the offset term, letting $(r_{j,t}^{\T} L)$ be the regression
vector with `coefficient' $\zeta$, treated as a
vector of `parameters'. The modal value $\zeta_j^{\ast}$ solves $\frac{\partial}{\partial \zeta}F_{j}(\zeta_j^{\ast})=0$ and can be obtained using
Newton-Raphson. At convergence we set
$\Sigma_j^{\ast}=\Sigma_j(\zeta^{\ast})$.
For $n_j$ sufficiently large, it can be shown that, for the canonical link $b$ which defines the density \eqref{Eq: ExpFamDensity},
$E[\Sigma_j(\zeta)$ for all $\zeta$
is positive definite and hence the Newton-Raphson method tends to converge to the modal
solution from any starting point. In practice this procedure has always converged from initiating value $\zeta_j^{(0)}=0$.

Noting that $\zeta_j^{\ast}(\Psi)$ and $\Sigma_j^{\ast}(\Psi)$ are functions of $\Psi$ the Laplace approximation to \eqref{Eq: Lj(Psi) Supplement} is 
\[
L_{j}^{(1)}(\Psi)=(\det\Sigma_{j}^{\ast}(\Psi))^{1/2}\exp(F_{j}(\zeta_{j}^{\ast}(\Psi)))
\]
giving the approximate log-likelihood for the $j$th state as
\begin{align*}
	l_{j}^{(1)}(\Psi)  =&\log\det(\Sigma_{j}^{\ast}(\Psi))^{1/2}-\frac{\zeta_{j}^{\ast}(\Psi)^{2}}{2}+\sum_{t=1}^{n}c(y_{jt})\\
	&+\sum_{t=1}^{n}\left[
	y_{jt}W_{jt}(\zeta_{j}^{\ast}(\Psi))-\exp(W_{jt}(\zeta_{j}^{\ast}(\Psi)))\right]
\end{align*}
which can be combined to give the overall approximate log likelihood as
\begin{equation}\label{Eq: LikeApproxLap}
	l^{(1)}(\Psi)=\sum_{j=1}^{J}l_{j}^{(1)}(\Psi)
\end{equation}
It is, of course, necessary to recompute the $J$ Laplace approximations for each value of $\Psi$ used in the iterative scheme used to optimize $l^{(1)}(\Psi)$ in \eqref{Eq: LikeApproxLap} 

The Laplace Approximation is also used to determine a basis for adaptive Gaussian
quadrature methods which have proved to be successful in improving the approximation to the likelihoods in non-linear and non-Gaussian mixed effects modelling and is implemented in a number of widely used software systems - see \cite{pinheiro1995approximations}
and \cite{pinheiro2006efficient}. The following description of adaptive Gaussian quadrature follows that of
\cite{pinheiro2006efficient}.
Let $\Sigma_{j}^{\ast}=K_{j}^{\ast}%
K_{j}^{\ast T}$ be the Cholesky decomposition of $\Sigma_{j}^{\ast}$.
Let $\zeta_{i},w_{j}$, $i=1,\dots,Q$ be the abscissas for the one
dimensional Gaussian quadrature rule with Q points based on the $\exp(-x^{2})$ kernel, let
$\zeta_{\mathcal{I}} = \zeta_{i_{1},\dots,i_{Q}} = (\zeta_{i_{1}}%
,\dots,\zeta_{i_{Q}})^{\T}  $ be a point in the $d$-dimensional grid of quadrature points where
$\mathcal{I} = (i_{1},\dots,i_{Q})$ and let
$\mathcal{W}_{\mathcal{I}} = \exp(\Vert\zeta_{\mathcal{I}} \Vert^{2})
\prod_{k=1}^{d} w_{i_{k}}$
be the quadrature weights assigned to $\zeta_{\mathcal{I}}$.

Using the $d^Q$ quadrature points and associated points, the $j$th integral in \eqref{Eq: loglik j in z} is approximated by
\begin{equation} \label{Eq: AGQ jth Likelihood}
	L_{j}^{(Q)}(\Psi)  =\frac{\det(K_{j}^{\ast})}{\pi^{d/2}}\sum_{\mathcal{I}}\exp\left[
	F_{j}(\zeta_{j}^{\ast}+K_{j}^{\ast}\sqrt{2}\zeta_{j}|\Psi)\right]
	\mathcal{W}_{\mathcal{I}}
\end{equation}
giving the approximation to the overall likelihood as
\begin{equation}
	l^{(Q)}(\Psi)
	=\sum_{j=1}^{J}\log L_{j}^{(Q)}(\Psi) \label{Eq: LikeApproxAGQ}
\end{equation}
When $Q=1$, $l^{(Q)}(\theta)$ is the Laplace approximation defined in (\ref{Eq: LikeApproxLap}).

However adaptive Gaussian quadrature, can provide
multipoint approximations to the integrals in \eqref{Eq: GLARMAlogLik jth 1stDeriv} and \eqref{Eq: GLARMAlogLik jth 2ndDeriv}. The same quadrature points $\zeta_{\mathcal{I}}$ used to in the approximations \eqref{Eq: AGQ jth Likelihood} are also be used for integrals in \eqref{Eq: GLARMAlogLik jth 1stDeriv} and \eqref{Eq: GLARMAlogLik jth 2ndDeriv}.

This gives the approximation to  $\dot{l}(\Psi)$ as 
\begin{align*}
	\dot{l}
	^{(Q)}(\Psi)=&\sum_{j=1}^{J}\frac
	{|L_{j}^{\ast}|}{L_{j}^{(Q)}(\Psi)\pi^{d/2}}\sum_{\mathbf{i}}\left(
	\frac{\partial}{\partial\Psi}l_{j}(\Psi|\zeta_{j}^{\ast}+K_{j}^{\ast}\sqrt
	{2}\zeta_{\mathbf{i}})\right) \\
	&\times \exp\left[  F_{j}(\zeta_{j}^{\ast}+K_{j}^{\ast
	}\sqrt{2}\zeta_{\mathbf{i}}|\Psi)\right]  \mathcal{W}_{\mathbf{i}}.
\end{align*}
with a similar expression for the approximation to the second derivative which we denote $\ddot{l}^{(Q)}(\Psi)$. 

\section{Appendix B: Accuracy and Execution Time Tradeoff}

Table \ref{Tab: TimeAccuracy} reports on the impact of increasing the number of quadrature points on the accuracy of estimation of the likelihood $l^{(Q)}(\Psi)$ defined in \eqref{Eq: LikeApproxAGQ} and standard errors (based on the inverse of the matrix of second derivatives estimated by $\ddot{l}^{(Q)}(\Psi)$ as well as the computation time required for one iteration of the Newton-Raphson optimization scheme when $\Psi^{(k)}$ is close to the optimizing value. Time trials are for the final model of the last section specified by \eqref{Eqn: Final Model Wjt} which has $3$ fixed effects parameters, $4$ parameters to specify the covariance for 3 random effects and $5$ serial dependence parameters. Thus there are $S=12$ parameters in $\Psi$ but because each of the $5$ serial dependence parameters is present in a subset of the $J$ cases, the total number of integrals to be estimated is $32 \times(8+1)^2=2592$ in total to estimate the overall likelihood and its first and second derivatives. Each of these integrals is based on $3^Q$ quadrature points. Note that when using $Q=5$ quadrature points in each dimension the time per iteration is slightly over $2$ minutes. We have found that using a smaller value of $Q$ in the initial stages of convergence and switching to a higher value of $Q$ for final refinement helps reduce overall computational time.


The log likelihood has essentially stabilised with $Q=4$ quadrature points in each of $d=3$ dimensions. The penultimate column of Table \ref{Tab: TimeAccuracy} shows the maximum percentage change in the parameter values as $Q$ is incremented by $1$ and the last column the maximum percentage change in the standard errors. For practical use there is little motivation for increasing the number of quadrature points beyond $Q=5$. Computation time per iteration for $Q=5$ was $2.19$ minutes per iteration.

\bibliography{DVCrossSecTSwithODmodelsarXiv}
\bibliographystyle{imsart-nameyear}
\newpage

\begin{table}
	\caption{\label{Tab: RE Estimates} Results from fitting random effects model \eqref{Eqn: Final Model Wjt} with GLARMA serial depenendence model to $32$ binary time series of positive responses from musical listeners. Parameters $L_{ij}$ are the lower triangular elements in $\Sigma=L L^\T$ in the covariance matrix for random effects on the Intercept, Cubic and Quadratic (in that order) orthogonal polynomial regressors}
	\begin{center}
		\begin{tabular}[c]{|l||l|r|r|}\hline
			Component & Parameter & Estimate & Stan. Err.\\\hline
			Fixed &Intercept & $-1.694$ & $0.140$\\\hline
			effects & Quadratic & $2.158$ & $0.359$\\\hline
			&Cubic     & $0.852$ & $ 0.626 $\\\hline
			Random &$L_{11}$ & $0.863$ & $0.110$\\\hline
			Effects &$L_{22}$ & $1.442$ & $0.367$\\\hline
			& $L_{33}$ & $2.379$ & $0.613$\\\hline
			&$L_{31}$ & $1.519$ & $0.648$\\\hline
			Serial  &$\phi_{1}$ (2 NM) & $0.364$ & $0.063$\\\hline
			Dependence& $\phi_{2}$ (1EA, 1M, 2NM)  & $0.473$ & $0.040$\\\hline
			&$\phi_{3}$ (2EA, 4M, 5NM)& $0.720$   & $0.016$\\\hline
			&$\phi_{4}$ (4EA, 3M, 3NM)& $0.776$  & $0.013$\\\hline
			&$\phi_{5}$ (1EA, 4NM)& $0.861$ & $0.013$\\\hline
		\end{tabular}
	\end{center}
\end{table}

\begin{table}
	\caption{\label{Tab: TimeAccuracy} Number of Quadrature Points: impact on Speed and Accuracy ($Q$ is the number of quadrature points in each of $d$ random effect dimensions, $T$ is the time taken in minutes for a single iteration of the Newton-Raphson parameter optimation, $\log L$ is the overall log likelihood. Final two columns show maximum percentage changes over parameter estimates and their standard errors respectively as $Q \to Q+1$. Timing using a 64-bit Windows 8 machine with 8GB of RAM running on an Intel Core i5-4300U, 1.9GHz/2.50GHz CPU}
	\begin{center}
		\begin{tabular}[c]{|r|r|r|r|r|r|}\hline
			$Q$  & $d^Q$ &  Time &     $\log L$ & Parameters&  \% S.E's\\ 
			& & & &\% change& \% change \\\hline
			2  &  9 & 0.291 & 9867.234   &     -  &     - \\
			3  &  27 & 0.596 & 9867.020   &   2.327 &   29.462\\
			4  &  81 & 1.196 & 9866.947   &   1.390  &   0.905\\
			5  & 243 & 2.193 & 9866.939   &   0.563  &   0.247\\
			6  & 729 & 4.260 & 9866.938   &   0.236   &  0.114\\
			7 & 2187 & 6.463 & 9866.938   &   0.092  &   0.052\\
			\hline
		\end{tabular}
		
	\end{center}
\end{table}

\end{document}